\documentclass[sn-nature]{sn-jnl}
\usepackage{graphicx}%
\usepackage{multirow}%
\usepackage{amsmath,amssymb,amsfonts}%
\usepackage{amsthm}%
\usepackage{cleveref}
\usepackage{mathrsfs}%
\usepackage[title]{appendix}%
\usepackage{xcolor}%
\usepackage{textcomp}%
\usepackage{manyfoot}%
\usepackage{booktabs}%
\usepackage{algorithm}%
\usepackage{algorithmicx}%
\usepackage{algpseudocode}%
\usepackage{listings}%
\usepackage{comment}
\usepackage{here}
\renewcommand{\figurename}{Fig.}
\renewcommand{\tablename}{Table}
\usepackage[labelsep=period]{caption}
\usepackage{xcolor}
\usepackage{booktabs}
\usepackage{float}
\usepackage{tikz} 
\usepackage{bm} 
\usepackage{verbatim}
\usepackage{multirow}
\usepackage{subcaption}
\usepackage{placeins}
\usepackage{pgfplots}
\theoremstyle{thmstyleone}%
\theoremstyle{thmstyletwo}%
\newtheorem{example}{Example}%
\theoremstyle{thmstylethree}%
\newtheorem{definition}{Definition}%
\begin{document}

\title{Community Quality and Influence Maximization: An Empirical Study }
\author*[1]{\fnm{Motaz} \sur{ Ben Hassine}}\email{benhassine@cril.fr}

\affil*[1]{\orgdiv{CRIL}, \orgname{University of Artois and CNRS}, \orgaddress{ \city{Lens}, \country{France}}}
\abstract{
Influence maximization in social networks plays a vital role in applications such as viral marketing, epidemiology, product recommendation, opinion mining, and counter-terrorism. A common approach identifies seed nodes by first detecting disjoint communities and subsequently selecting representative nodes from these communities. However, whether the quality of detected communities consistently affects the spread of influence under the Independent Cascade model remains unclear. This paper addresses this question by extending a previously proposed disjoint community detection method, termed \textit{$\alpha$-Hierarchical Clustering}, to the influence maximization problem under the Independent Cascade model. The proposed method is compared with an alternative approach that employs the same seed selection criteria but relies on communities of lower quality obtained through standard Hierarchical Clustering. The former is referred to as \textit{Hierarchical Clustering-based Influence Maximization}, while the latter, which leverages higher-quality community structures to guide seed selection, is termed \textit{$\alpha$-Hierarchical Clustering-based Influence Maximization}. Extensive experiments are performed on multiple real-world datasets to assess the effectiveness of both methods. The results demonstrate that higher-quality community structures substantially improve information diffusion under the Independent Cascade model, particularly when the propagation probability is low. These findings underscore the critical importance of community quality in guiding effective seed selection for influence maximization in complex networks.
}
\maketitle
\section{Introduction}\label{sec:sec1}
Influence in social networks has gained significant attention in recent years \cite{fratiglioni2000influence}. This growing interest has motivated extensive research aimed at understanding how influence propagates, including the identification of key structural properties such as opinion leaders and communities of interest. To address these challenges, researchers have applied various data mining techniques, including community detection and \textbf{Influence Maximization} (IM).
The concept of IM was first introduced by Domingos et al. \cite{domingos2001mining} in the context of viral marketing. Later, Kempe et al. \cite{kempe2003maximizing} formalized IM as an optimization problem. Given a positive integer $k$, the goal is to identify a seed set $S$ of $k$ influential nodes that maximizes the expected spread of influence. Kempe et al. also demonstrated that the IM problem is $\mathcal{NP}$-hard and proposed two widely adopted diffusion models: the Linear Threshold Model (LTM) and the Independent Cascade Model (IC).
In the IC model, influence spreads over a directed graph $G = (V, E)$ where each edge $(u,v)$ is associated with an activation probability $p_{uv}$. Starting from an initial seed set $S$, the diffusion unfolds in discrete time steps. At each step $t_i$ ($i \geq 0$), newly activated nodes are each given one chance to independently activate their currently inactive neighbors. A node $u$ attempts to activate a neighbor $v$ with probability $p_{uv}$; if successful, $v$ becomes active at time $t_{i+1}$. Regardless of success or failure, $u$ cannot attempt to activate $v$ again. The process terminates when no more activations are possible \cite{wang2018modeling}.
Despite its importance, IM remains challenging due to both computational complexity and the difficulty of selecting high-quality seed nodes. Kempe et al. \cite{kempe2003maximizing} proposed the first Greedy algorithm for IM, which iteratively adds the node that provides the largest marginal gain in influence spread. They proved that this algorithm achieves a $(1 - \frac{1}{e}) \approx 0.63$ approximation of the optimal solution. However, the Greedy method is computationally demanding because it repeatedly simulates the diffusion process.
To improve efficiency, Leskovec et al. \cite{leskovec2007cost} introduced the Cost-Effective Lazy Forward (CELF) algorithm, which accelerates the Greedy procedure by a factor of about 700. Subsequently, Goyal et al. \cite{goyal2011celf++} proposed CELF++, which further improves runtime performance. Borgs et al. \cite{borgs2014maximizing} then introduced the Reverse Influence Sampling (RIS) framework, providing a more scalable alternative to Greedy-based methods.
Beyond these approaches, several IM methods rely on community detection. In social networks, a community is a group of users who share common interests or characteristics. In graph theory, it refers to a set of vertices that are densely connected internally but sparsely connected to the rest of the graph. Community-based IM methods exploit these structural properties to guide seed selection. For example, Bouyer et al. \cite{bouyer2023fip} proposed a method using overlapping communities to select influential overlapping nodes. Chen et al. \cite{chen2014cim} introduced the Community-based Influence Maximization (CIM) framework, which uses Hierarchical Clustering (HC) under  the Heat Diffusion Model (HDM) \cite{ma2008mining}. Numerous recent influence maximization approaches rely on community detection as a foundation, including \cite{kazemzadeh2022influence,tyagi2024influence,devi2025community,ahmad2025learning}.
Recently, I improved the HC method by proposing a new similarity measure called $\alpha$-Structural Similarity ($\alpha$-2S), which accounts for both common neighbors and the interconnection between them. This led to the development of the $\alpha$-Hierarchical Clustering ($\alpha$-HC) algorithm, which replaces the traditional Structural Similarity (2S) measure with $\alpha$-2S \cite{hassine2023non}.
However, it remained unclear whether the higher-quality community structures produced by $\alpha$-HC consistently affect the spread of influence and lead to more effective seed selection under the Independent Cascade (IC) model \cite{kempe2003maximizing}. To address this question, I extend the $\alpha$-HC framework to the influence maximization problem and introduce two methods for comparison: \textit{Hierarchical Clustering-based Influence Maximization} (HCIM), which relies on standard HC with lower-quality communities, and \textit{$\alpha$-Hierarchical Clustering-based Influence Maximization} ($\alpha$-HCIM), which leverages the higher-quality community structures detected by $\alpha$-HC to guide seed selection.

The remainder of the paper is structured as follows.  
Section~\ref{sec:sec2} introduces the formal notations.  
Section~\ref{sec:sec3} presents the proposed $\alpha$-HCIM approach.  
Section~\ref{sec:sec4} describes my experiments on real-world social networks.  
Finally, Section~\ref{sec:sec5} concludes the paper and outlines future research directions.

\section{Formal notations}\label{sec:sec2}
%
%
Consider an undirected graph $G = (V,E)$, where $V$ is the  set of vertices, and $E$ is the set of edges.
The set of \textbf{neighbors} of a node $u \in V$ is defined as $Nei(u) = \{v ~ | ~ (u,v)\in E\}$.
The \textbf{degree} of $u\in V$ is then $|Nei(u)|$.
 $G$ can  be divided into numerous subgroups called \textbf{communities}, denoted by 
$C_{G} = \{c_1,c_2,\ldots,c_m\}$.  
Let $d_{\max}$ be the \textbf{maximum degree} of $G$, i.e,
$d_{\max} = \max_{|Nei(u)|}\{u\in V\}$. Given a set of nodes $X\subseteq V$,  a \textbf{subgraph} induced by $X$, denoted as $G_X=(X,E_X)$, 
is a graph over $X$ s.t. $E_X = \{(u,v)\in E ~|~ u,v \in X\}$.
Besides, let $d_{av}$  be the \textbf{average number of neighbors} of all the vertices in $X$, i.e., $d_{av}=\frac{1}{|X|} \sum_{u \in X}|Nei(u)|$. 
Let $\phi(S)$ represent \textbf{the final active set } produced by the stochastic diffusion model, specifically the IC model. Additionally, let denote $\sigma(S) = E(|\phi(S)|)$ as \textbf{the influence spread } of set $S$.
This indicates  the  average size of the final active sets the expected size of the final active sets. Let consider $L$ be the path between the nodes $u$ and $v$, then,  $d(u,v)$ is called  the \textbf{distance} between   $u$ and $v$ defined as $d(u,v)=\min_{m-1}\{L=(v_1,v_2, \ldots, v_m)~ s.t.~ v_1=u,~  v_m=v,~ \forall ~ 1\leq i <m, ~ (v_i,v_{i+1}) \in E \} $. \textbf{The coverage} of a node $u \in V$  is defined as  $N^{\theta}(u)= \{ v \in V ~ | ~ d(u,v) \leq \theta  \} \cup \{u\}$. Further, denote $sizeCom(u)= |c| $  where $u \in c$ and  $c \in C_G$ the size of the community which $u$ belongs to it. A node $u$ is called an \textbf{overlapping node} if 
$\exists$  $c_i, c_j \in C_G$, $1\leq i < j \leq m$ s.t. 
$c_i \cap Nei(u) \neq \emptyset$ and  
$c_j \cap Nei(u) \neq \emptyset$.

\begin{definition}[Influence Maximization problem \cite{kempe2003maximizing}] \label{defIM}
Given an undirected graph $G=(V,E)$  and $k>0$. Find a seed set $S \subseteq V$ where  $|S| = k$  and $k \ll |V|$ s.t.  $\forall \; S' \subseteq V$  where $|S'|=k$,  I have  $\sigma(S) \geq \sigma(S’)$.
\end{definition}

\section{An Influence maximization approach based on $\alpha$-HC }\label{sec:sec3}

%

In this section,  $\alpha$-HC method  \cite{hassine2023non} is extended for the influence maximization problem, referred to $\alpha$-HCIM. %
Overlapping nodes can be added in the seed set  because they can link several communities in such a way that it contributes to the  effective propagation of information. However, not all overlapping nodes have the same importance and then I propose a novel score called \textbf{propagator-score}  which is described in the definition that follows.
\\
\begin{definition}[propagator-score]\label{defpropscore}
 Let $G= (V,E)$ be an undirected graph. Let  $v \in V$ be an overlapping node, and $G_{N^{\theta}(v)} = (N^{\theta}(v),E_{N^{\theta}(v)})$ the subgraph induced by $N^{\theta}(v)$ then the propagator-score noted $score(v)$ is defined as :
 \begin{equation}\label{defpropscore}
     score(v) = \sum_{(u,w) \; \in \; E_{N^{\theta}(v)}}|Nei(u) \cap Nei(w)|
 \end{equation}
\end{definition}
\begin{example} \label{exp3}
Consider the undirected graph shown in Figure~\ref{figaa}.
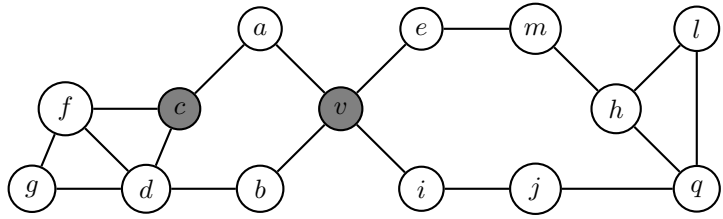
\begin{figure}[H]
 \captionsetup{labelfont=bf}
\begin{center}
\begin{tikzpicture}[node distance={15mm}, thick, main/.style = {draw, circle}] 
\node[main] (a) {$a$}; 
\node[main, fill=gray] (c) [ below left  of=a] {$c$};
\node[main] (f) [ left  of=c] {$f$};
\node[main, fill=gray] (v) [ below right  of=a]  {$v$};
\node[main] (e) [ above right  of=v]  {$e$};
\node[main] (m) [ right  of=e]  {$m$};
\node[main] (h) [ below right  of=m]  {$h$};
\node[main] (i) [ below right  of=v]  {$i$};
\node[main] (l) [ above right  of=h]  {$l$};
\node[main] (j) [ right  of=i]  {$j$};
\node[main] (q) [ below right  of=h]  {$q$};
\node[main] (b) [below left  of=v]  {$b$}; 
\node[main] (d) [ left  of=b] {$d$};
\node[main] (g) [ left  of=d]  {$g$};
\draw (a) -- (c);
\draw (a) -- (v);
\draw (b) -- (v);
\draw (e) -- (v);
\draw (e) -- (m);
\draw (i) -- (v);
\draw (h) -- (m);
\draw (h) -- (l);
\draw (h) -- (q);
\draw (i) -- (j);
\draw (b) -- (d);
\draw (c) -- (d);
\draw (f) -- (c);
\draw (f) -- (d);
\draw (f) -- (g);
\draw (g) -- (d);
\draw (j) -- (q);
\draw (l) -- (q);
\end{tikzpicture}
%
\end{center}
Supposing that $v$ and $c$  are overlapping nodes and $\theta=2$, then:
{\scriptsize
\begin{align*}
    N^{2}(v) &= \{v,a,e,i,b,m,j,c,d\}\\
    N^{2}(c) &= \{c,a,v,f,d,b,g\}
\end{align*}
}
then : 
{\scriptsize
\begin{align*}
    score(v) &= 0\\
    score(c) &= 5\\
\end{align*}
}
\caption{Example for calculating the score of the nodes $v$ and $c$ with $\theta=2$.} \label{figaa}
\end{figure}
\end{example}
In example \ref{exp3}, based on the scores,  node $v$ appears to be more important  than  node $c$ for selection in the seed set. This is because the score calculates the sum of common neighbors for each pair of nodes linked by an edge in the subgraph induced by $N^2(v)$, which plays a role in mitigating conflicts during the propagation of information. The node with the minimum score is selected to be a candidate in the seed set. When this sum is minimized, it helps to avoid these conflicts. 
Indeed, let  assume that node $c$ is selected in the seed set. Consider a scenario where both nodes $f$ and $d$ simultaneously receive information from  node $c$ at time $t$. In this situation, an information propagation conflict arises because node $g$ will subsequently receive information from both nodes $f$ and $d$ at time $t+1$. These conflicts result in a reduction in the number of impacted nodes that successfully receive the information.
In the proposed approach, I begin with a set of communities called $C_G$. I then perform a decomposition into two new sets. The first set consists of communities with a single node, while the second set contains the remaining communities.
\begin{algorithm}[H]
\caption{$\alpha$-HCIM}\label{alg:two}
\begin{algorithmic}[1]
\Require $G(V,E), \alpha, k $
\Ensure $S$
\State $C_G \Leftarrow  $ $\alpha$-HC
\State $S \Leftarrow \emptyset$
\State $C_o \Leftarrow  \{ \forall\; c_o \in C_G ~ | ~ |c_o|=1\}$
\State $C_b \Leftarrow  \{ \forall\; c_b \in C_G ~ | ~ |c_b|>1\}$
\While{$|S|< k $ and $C_b \neq \emptyset $}
\For{$ c_b \in C_b$}
\State $no \Leftarrow \underset{ u \in c_b }{\arg\max} ~ \sigma(\{u\}) $
\State $S \Leftarrow S \cup no$
\State $c_b \Leftarrow c_b \setminus S $
\EndFor
\EndWhile
\While{$|S|< k $}
 \State $n \Leftarrow \underset{ u_o \in c_o,~ c_o \in C_o  }{\arg\min}  score(u_o) $
 \State $S \Leftarrow S \cup \{n\}$
 \State $c_o \Leftarrow c_o \setminus S $
\EndWhile
\While{ True }
\State $n_1 \Leftarrow \underset{u_o \in c_o,~ c_o \in C_o}{\arg\min}  score(u_o)$
\State $spread_1 \Leftarrow \sigma(S)$
\State $n_2 \Leftarrow \underset{u \in S}{\arg\min}~~sizeCom(u) $
\State $S\Leftarrow S \setminus \{n_2\}$
\State $S \Leftarrow S \cup \{n_1\} $
\State $spread_2 \Leftarrow \sigma(S)$
\If{$spread_2 >spread_1 $ or $C_o = \emptyset $ }
\State \textbf{break}
\EndIf
\State $S\Leftarrow S \setminus \{n_1\}$
\State $S \Leftarrow S \cup \{n_2\} $
\State $C_o \Leftarrow C_o \setminus \{n_1\}$
\EndWhile
\State \textbf{return} $S$
\end{algorithmic}
\end{algorithm}
Subsequently, I apply a sorting operation to the second set, arranging the communities in descending order based on their respective sizes. Significantly, larger communities are of greater importance due to their critical role in information propagation. Following this sorting process, for each community within the second set, I select the node that maximizes propagation within the subgraph induced by the community set under the IC model. If the size of the seed set remains below $k$, I then select the remaining nodes from the set of single-node communities using the propagator score. 
A node is chosen if it has the minimum value of the propagator score. After this selection process, I make a small enhancement to the seed set. Specifically, I compute the spreading rate of the seed set. Subsequently, I remove the last node added to the seed set. This node is either selected from the smallest community or the set of single-node communities or selected from larger communities but with less expected information spread value, compared with the nodes selected before it. The removed node is replaced with a node selected from the set of single-node communities. I then recompute the seed set. If the spreading rate is higher than before, I retain the current seed set as the final result. If not, I search for an alternative seed set. It should be noted that the HCIM approach follows exactly the same methodology as $\alpha$-HCIM, which is formally presented in Algorithm \ref{alg:two}, with the sole difference being that communities are detected using the standard HC method instead of $\alpha$-HC. In other words, for HCIM, $C_G \Leftarrow$ HC, whereas for $\alpha$-HCIM, $C_G \Leftarrow$ $\alpha$-HC.
%
%
%
%

\paragraph{\textbf{Complexity analysis.}}  The $\alpha$-HC   requires $O(|V|d_{max}^3 + 4|E||X|d_{av} + (|E|+|V|)|V|)$ \cite{hassine2023non}. Let consider $r$ as the number of Monte Carlo simulations then the IC model takes $O(rk|E|)$ time to estimate the influence spread \cite{trivedi2020efficient} \cite{ye2022influence}. Besides, when given a set $X \subset V$, extracting the subgraph induced by $X$ requires $O(|X|^2)$ \cite{induced}. Moreover, for the first while loop, let  assume that $|S| = k_1$. Let consider $B \subset N^{2}(u_o) $ and $D \subset N^{2}(u_o)$ such that $B\neq D$. Furthermore, let  consider $m_b$ as the subgraph's edges number induced by $c_b$. Finally, let  denote a community $c \in C_G$. Then, $\alpha-$HCIM requires $O(F)$ where $F= (k_1 |C_b| (|c_b|^2 + |c_b| r m_b)) + (k - k_1)|C_o|(|N^{2}(u_o)|^2 + |E_{N^{2}(u_o)}| min(|B|+|D|) ) +  |C_o|(|C_o|(|N^{2}(u_o)|^2 + |E_{N^{2}(u_o)}| min(|B|+|D|) ) + 2 k r |E| + k|c||C_G|)$. The complexity is polynomial and  is moderately high, which indicates that IM is an $\mathcal{NP}$-hard problem that can always be improved on the runtime side. 
\section {Experiments} \label{sec:sec4}
This section presents the experiments the influence maximization. Table \ref{tab1}  illustrates the datasets utilized.
\begin{table}[h!]
\caption{Real-world networks.} \label{tab1}
\begin{tabular}{{@{}lll@{}}}
\toprule
\textbf{Datasets} & \textbf{Nodes/Edges}   & \textbf{Source}\\
\midrule
Karate & 34/78    & \cite{zachary1977information}\\
Dolphin & 62/159  & \cite{lusseau2003bottlenose} \\
Books & 105/441  & \cite{krebs2004books}\\
Email-Eu-Core &  1005/25571  & \cite{leskovec2014snap} \\
Facebook-artist & 50515/819306   &\cite{rozemberczki2019gemsec}\\
\bottomrule
\end{tabular}
%
%
\end{table}

To answer my question, I performed two kinds of experiments. \footnote{The source code is available at: \url{https://github.com/2x254/Influence-Maximization}}
In the first phase of the experiment, I compared $\alpha$-HCIM and HCIM, with $\alpha$  set to 1. Both methods employed the same criteria for selecting seed nodes, but HCIM relied on Hierarchical Clustering (HC) \cite{chen2014cim}. I tested these approaches on two small-world datasets, namely the Karate network and the Dolphin network. I varied the probability parameter $p$ of  the IC model, from $0.1$ to $0.9$ while keeping $k$ constant  where $k \ll |V|$ (specifically, $k=4$). It should be noted that  the number of Monte Carlo simulations for IC model was set to 100 ($r=100$). For experimental reasons, I fixed $\theta=2$. Additionally, I performed experiments by fixing the value of $p$ (specifically, $p=0.1$) and varying $k$ from $4$ to $10$.
In the second phase of the experiment, I set the value of $p$ to the lowest probability ($p=0.1$). My goal was to evaluate the performance of these approaches under the most challenging scenarios of the IC diffusion model, where information propagates very slowly. I compared $\alpha$-HCIM and HCIM with other datasets. I selected various networks based on their sizes, particularly focusing on the number of edges. In my experiments, I opted for three small-world datasets (Karate, Books, and Dolphin), one moderate dataset (Email-Eu-Core), and one large dataset (Facebook-artist). The selection of Facebook-artist as the large dataset was primarily due to the computational demands of Monte Carlo simulations. To ensure fairness in evaluation across datasets with different sizes, I set $k=10$. I also compared $\alpha$-HCIM with two other well-known approaches  called CELF \cite{leskovec2007cost} and Greedy \cite{kempe2003maximizing} known for achieving high information spread for an approximation near to $1 - \frac{1}{e}$  compared with the optimal solution. My comparison was based on  the expected information spread and running time.
These experiments were conducted on a machine with a 4-Core E5-2637 CPU (3.5GHz) and 128GB of RAM. It should be noted  that the maximum allowable running time delay was 168 hours.

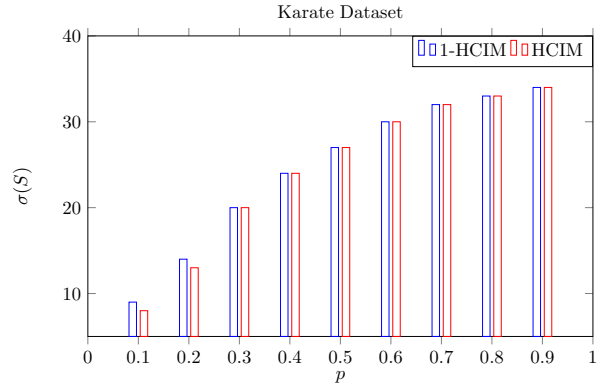
\begin{figure}[H] 
\centering
\captionsetup{labelfont=bf}

     \scalebox{0.7}{
         \begin{tikzpicture}
\begin{axis}[
    	title= Karate Dataset,
     ybar,
     bar width=4,
     x=270pt,
     legend style={
    legend columns=4,
    at={(1,1)}},
	ylabel=$\sigma(S)$,
    xlabel=$p$,
    xmin=0,
    xmax=1,
    ymin=5,
    ymax=40
]
\addplot [blue] coordinates {(0.1,9)
(0.2,14)
(0.3,20)
(0.4,24)
(0.5,27)
(0.6,30)
(0.7,32)
(0.8,33)
(0.9,34)
};
\addplot [red] coordinates {(0.1,8)
(0.2,13)
(0.3,20)
(0.4,24)
(0.5,27)
(0.6,30)
(0.7,32)
(0.8,33)
(0.9,34)
};
\legend{$1$-HCIM,HCIM}
\end{axis}
\end{tikzpicture}
}
\caption{$1$-HCIM  vs HCIM on Karate dataset based on expected information spread with  k =4.} \label{fig12}
\end{figure}

Firstly, I compared $1$-HCIM to HCIM in terms of expected information spread and running time. I selected the Karate and the Dolphin networks because $1$-HCIM performed better than HCIM in the phase of community detection. I aimed to determine  whether  the community quality could enhance information spread under the same selection criteria and using the same diffusion model, specifically the IC model. Figures \ref{fig12}, \ref{fig13}, \ref{fig14}, \ref{fig15}, \ref{fig16}, \ref{fig17}, \ref{fig18}, and \ref{fig19} illustrate the obtained results.
%
%
\begin{figure}[H] 
\centering
\captionsetup{labelfont=bf}
%

     \scalebox{0.7}{
         \begin{tikzpicture}
\begin{axis}[
    	title= Karate Dataset,
     ybar,
     bar width=4,
     x=270pt,
     legend style={
    legend columns=4,
    at={(1,1)}},
	ylabel=Running time(s),
    xlabel=$p$,
    xmin=0,
    xmax=1,
    ymin=0,
    ymax=5
]
\addplot [blue] coordinates {(0.1,0.3609)
(0.2,0.7153)
(0.3,1.3831)
(0.4,2.4977)
(0.5,2.3221)
(0.6,2.8484)
(0.7,3.2769)
(0.8,3.6855)
(0.9,3.83)
};
\addplot [red] coordinates {
(0.1,0.28)
(0.2,0.53)
(0.3,0.92)
(0.4,1.57)
(0.5,1.48)
(0.6,1.67)
(0.7,1.97)
(0.8,2.02)
(0.9,2.1)
};
\legend{$1$-HCIM,HCIM}
\end{axis}
\end{tikzpicture}
}
%
\caption{$1$-HCIM  vs HCIM on Karate dataset based on running time in seconds with k=4.} \label{fig13}
\end{figure}
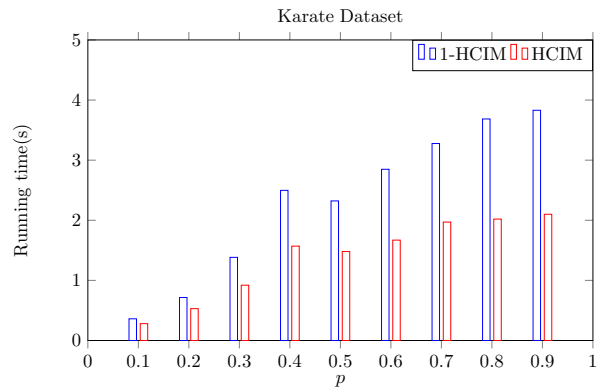

%
%
\begin{figure}[H] 
\centering
\captionsetup{labelfont=bf}
     \scalebox{0.7}{
         \begin{tikzpicture}
\begin{axis}[
    	title= Dolphin Dataset,
     ybar,
     bar width=4,
     x=270pt,
     legend style={
    legend columns=4,
    at={(1,1)}},
	ylabel=$\sigma(S)$,
    xlabel=$p$,
    xmin=0,
    xmax=1,
    ymin=0,
    ymax=70
]
\addplot [blue] coordinates {
(0.1,11)
(0.2,20)
(0.3,32)
(0.4,40)
(0.5,47)
(0.6,53)
(0.7,56)
(0.8,59)
(0.9,61)
};
\addplot [red] coordinates {
(0.1,10)
(0.2,20)
(0.3,32)
(0.4,41)
(0.5,47)
(0.6,53)
(0.7,56)
(0.8,59)
(0.9,61)
};
\legend{$1$-HCIM,HCIM}
\end{axis}
\end{tikzpicture}
}
\caption{$1$-HCIM  vs HCIM on Dolphins dataset based on expected information spread with  k =4.} \label{fig14}
\end{figure}
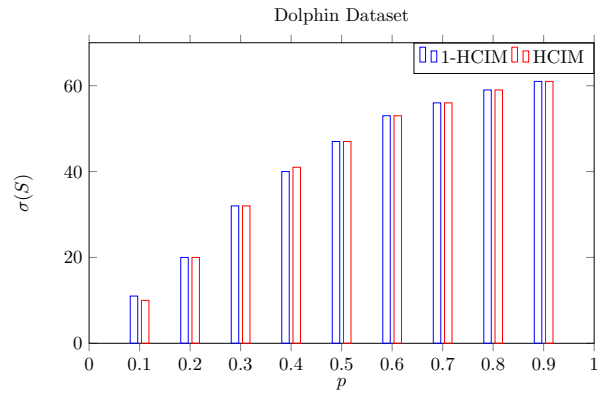
%
%
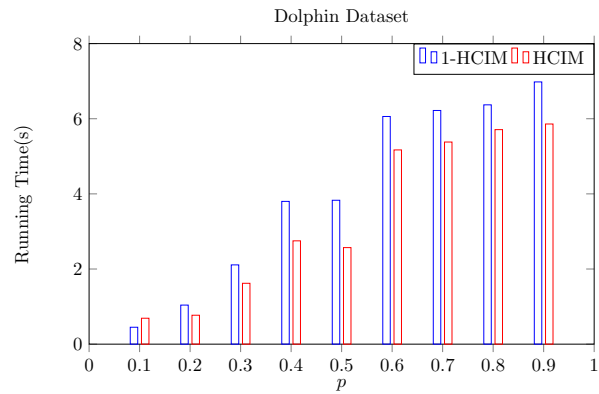
\begin{figure}[H] 
\centering
\captionsetup{labelfont=bf}
     \scalebox{0.7}{
         \begin{tikzpicture}
\begin{axis}[
    	title= Dolphin Dataset,
     ybar,
     bar width=4,
     x=270pt,
     legend style={
    legend columns=4,
    at={(1,1)}},
	ylabel=Running Time(s),
    xlabel=$p$,
    xmin=0,
    xmax=1,
    ymin=0,
    ymax=8
]
\addplot [blue] coordinates {
(0.1,0.45)
(0.2,1.04)
(0.3,2.11)
(0.4,3.8)
(0.5,3.83)
(0.6,6.06)
(0.7,6.22)
(0.8,6.37)
(0.9,6.98)
};
\addplot [red] coordinates {
(0.1,0.69)
(0.2,0.77)
(0.3,1.62)
(0.4,2.75)
(0.5,2.57)
(0.6,5.17)
(0.7,5.38)
(0.8,5.71)
(0.9,5.86)
};
\legend{$1$-HCIM,HCIM}
\end{axis}
\end{tikzpicture}
}
\caption{$1$-HCIM vs. HCIM on Dolphin dataset based on  running time in seconds with $k=4$.} \label{fig15}
\end{figure}
%
%
%
%
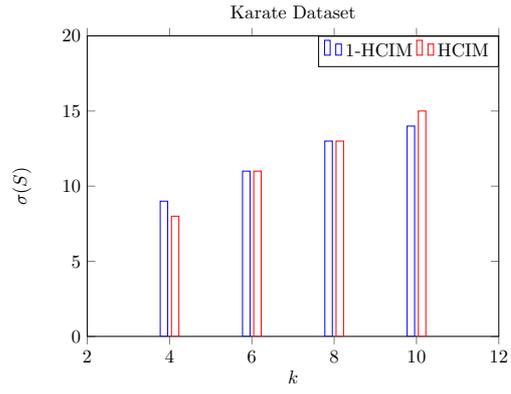
\begin{figure}[H] 
\centering
\captionsetup{labelfont=bf}
     \scalebox{0.7}{
         \begin{tikzpicture}
\begin{axis}[
    	title= Karate Dataset,
     ybar,
     bar width=4,
     x=22pt,
     legend style={
    legend columns=4,
    at={(1,1)}},
	ylabel=$\sigma(S)$,
    xlabel=$k$,
    xmin=2,
    xmax=12,
    xtick distance =2,
    ymin=0,
    ymax=20
]
\addplot [blue] coordinates {
(4,9)
(6,11)
(8,13)
(10,14)
};
\addplot [red] coordinates {
(4,8)
(6,11)
(8,13)
(10,15)
};
\legend{$1$-HCIM,HCIM}
\end{axis}
\end{tikzpicture}
}
\caption{$1$-HCIM  vs HCIM on Karate dataset based on expected information spread  with $p=0.1$.} \label{fig16}
\end{figure}
%
%
%
\begin{figure}[H] 
\centering
\captionsetup{labelfont=bf}
     \scalebox{0.7}{
         \begin{tikzpicture}
\begin{axis}[
    	title= Karate Dataset,
     ybar,
     bar width=4,
     x=22pt,
     legend style={
    legend columns=4,
    at={(1,1)}},
	ylabel=Running time(s),
    xlabel=$k$,
    xmin=2,
    xmax=12,
    xtick distance=2,
    ymin=0,
    ymax=1
]
\addplot [blue] coordinates {
(4,0.36)
(6,0.37)
(8,0.48)
(10,0.6)
};
\addplot [red] coordinates {
(4,0.28)
(6,0.43)
(8,0.43)
(10,0.4)
};
\legend{$1$-HCIM,HCIM}
\end{axis}
\end{tikzpicture}
}
\caption{$1$-HCIM  vs HCIM on Karate dataset based on running time in seconds  with $p=0.1$.} \label{fig17}
\end{figure}
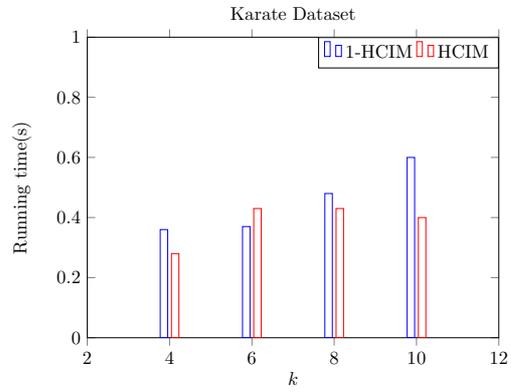
%
%
%
%
\begin{figure}[H] 
\centering
\captionsetup{labelfont=bf}
     \scalebox{0.7}{
         \begin{tikzpicture}
\begin{axis}[
    	title= Dolphin Dataset,
     ybar,
     bar width=4,
     x=22pt,
     legend style={
    legend columns=4,
    at={(1,1)}},
	ylabel=$\sigma(S)$,
    xlabel=$k$,
    xmin=2,
    xmax=12,
    xtick distance=2,
    ymin=0,
    ymax=25
]
\addplot [blue] coordinates {
(4,11)
(6,13)
(8,16)
(10,19)

};
\addplot [red] coordinates {
(4,10)
(6,13)
(8,15)
(10,18)
};
\legend{$1$-HCIM,HCIM}
\end{axis}
\end{tikzpicture}
}
\caption{$1$-HCIM  vs HCIM on Dolphins dataset based on expected information spread  with $p=0.1$.} \label{fig18}
\end{figure}
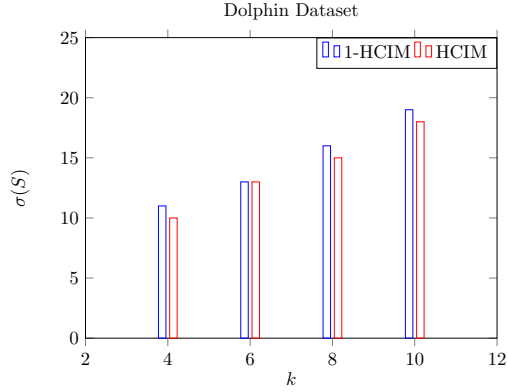
%
%
%
\begin{figure}[H] 
\centering
\captionsetup{labelfont=bf}
     \scalebox{0.7}{
         \begin{tikzpicture}
\begin{axis}[
    	title= Dolphin Dataset,
     ybar,
     bar width=4,
     x=22pt,
     legend style={
    legend columns=4,
    at={(1,1)}},
	ylabel=Running time(s),
    xlabel=$k$,
    xmin=2,
    xmax=12,
    xtick distance=2,
    ymin=0,
    ymax=1.5
]
\addplot [blue] coordinates {
(4,0.45)
(6,0.9)
(8,1.03)
(10,0.8)
};
\addplot [red] coordinates {
(4,0.69)
(6,0.39)
(8,0.89)
(10,1.07)
};
\legend{$1$-HCIM,HCIM}
\end{axis}
\end{tikzpicture}
}
\caption{$1$-HCIM  vs HCIM on Dolphins dataset based on running time in seconds  with $p=0.1$. } \label{fig19}
\end{figure}
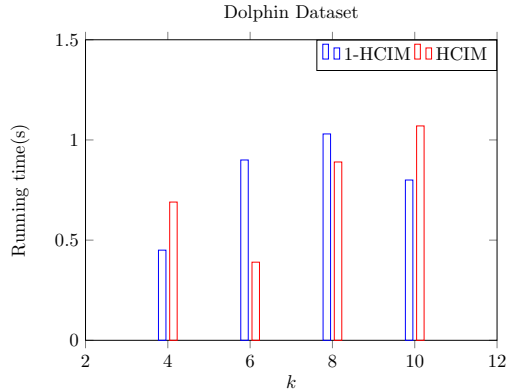
On the information spreading side, histograms reveal that for $k = 4$ and $p \leq 0.3$, 1-HCIM exhibits a better information spread compared to HCIM, or at least, it matches HCIM's performance. This finding holds particular interest in the context of the IC model, where information propagation is notably slow. 
However, if $p > 0.3$ or $k \neq 4$, 1-HCIM performs similarly overall, occasionally displaying slight improvements or setbacks. On the running time side, HCIM generally outperforms 1-HCIM due to its use of the 1-HC community detection approach, which comes with higher time complexity, particularly the $1$-2S similarity. 

Despite the significance of running time considerations, my primary focus in this paper is to address the question of whether the enhancement in information spread is contingent on the quality of the identified communities. To gain further insight into this question, I conducted additional comparisons on different datasets during the second phase of the experiment. 
Secondly, I maintained $p=0.1$ and I set $k=10$. I conducted an initial comparison between $1$-HCIM and HCIM. On the aspect of expected information spread, table \ref{tab8} demonstrates that $1$-HCIM outperforms HCIM on the Books network, Dolphin network, and Facebook-artist network with differences of 1, 1, and 888 active nodes, respectively. However, HCIM performs better on the Email network and the Karate network, with differences of 1 and 1, respectively.
Based on the results obtained from  the first and second phases of the experiment, I can infer that enhancing the community quality can improve information spread, especially when $p$  has a lower value within the IC model. However, it's important to note that this improvement may not be consistent across all scenarios.
Regarding the running time, HCIM generally outperforms $1$-HCIM, but there are instances where $1$-HCIM  is faster than HCIM, such as on the Dolphin and Email networks.

Furthermore, I conducted a second comparison between $1$-HCIM, CELF, and the GREEDY approach, and the results are presented in Tables \ref{tab9} and \ref{tab10}. 
In terms of running time, 1-HCIM outperforms all other approaches across all datasets. However, regarding  information spread, 1-HCIM exhibits a minor setback with only a modest difference. On the other hand, CELF and GREEDY face scalability issues, as demonstrated by exceeding the timeout during the execution on the Facebook-artist dataset without producing results.
\begin{table}[h!]
      \caption{$1$-HCIM vs. HCIM based on expected information spread and running time(s) with $k=10$ and $p=0.1$.}\label{tab8}
       \begin{tabular*}{\textwidth}{@{\extracolsep\fill}ccccc}
        \toprule
        \multirow{2}{*}{\textbf{Datasets}} & \multicolumn{2}{c}{\textbf{$1$-HCIM}} & \multicolumn{2}{c}{\textbf{HCIM}} \\
        \cmidrule{2-3} \cmidrule{4-5} & \textbf{Running time(s)} & \bm{$\sigma(S)$} & \textbf{Running time (s)} & \bm{$\sigma(S)$} \\
        \midrule
        Karate & 0.6 & 14 & \textbf{0.4} & \textbf{15} \\
        Books & 4.4 & \textbf{32} & \textbf{2.73} & 31 \\
        Dolphin & \textbf{0.8} & \textbf{19} & 1.07 & 18 \\
        Email-Eu-Core & \textbf{551.89} & 694 & 637.15 & \textbf{695} \\
        Facebook-artist & 24665.26 & \textbf{30633} & \textbf{4484.54} & 29745 \\
        \botrule
    \end{tabular*}
%
\end{table}
\begin{table}[h!]
      \caption{$1$-HCIM vs.CELF vs. GREEDY based  on expected information spread with $k=10$ and $p=0.1$.}\label{tab9}
\begin{tabular*}{\textwidth}{@{\extracolsep\fill}cccc}
\toprule
\multicolumn{4}{c}{\textbf{Comparison based on $\sigma(S)$}}\\
\midrule
\textbf{Datasets} & \textbf{$1$-HCIM} & \textbf{CELF}&  \textbf{GREEDY} \\
\midrule
Karate & 14 & \textbf{15} & \textbf{15} \\
Books  & 32&	\textbf{34} & \textbf{34} \\
Dolphin &\textbf{19} &18 & \textbf{19} \\ 
Email-Eu-Core & 694 & \textbf{702} & \textbf{702} \\
Facebook-artist  & \textbf{30633} &	- &	- \\
\botrule
\end{tabular*}
\end{table}
%
\begin{table}[h!]
      \caption{$1$-HCIM vs.CELF vs. GREEDY based  on running time(s) with $k=10$ and $p=0.1$.}\label{tab10}
\begin{tabular*}{\textwidth}{@{\extracolsep\fill}cccc}
\toprule
\multicolumn{4}{c}{\textbf{Comparison based on running time(s)}}\\
\midrule
\textbf{Datasets} & \textbf{$1$-HCIM} & \textbf{CELF}&  \textbf{GREEDY} \\
\midrule
Karate & \textbf{0.6} & 0.99  &  6.23 \\
Books  & \textbf{4.4} &	 10.31 & 64.43 \\
Dolphin & \textbf{0.8} &	 2.61 &	16.01  \\
Email-Eu-Core & \textbf{551.89} & 3419.6  & 18402,01  \\
Facebook-artist  & \textbf{24665.26} & -  &	- \\
\botrule
\end{tabular*}
\end{table}
\section {Conclusion} \label{sec:sec5}
In summary, in this paper
In this work, I extended the $\alpha$-HC approach to the influence maximization problem under the Independent Cascade (IC) model, resulting in the proposed framework $\alpha$-HCIM. Experimental results show that, particularly when the IC model operates with a low propagation probability $p$, higher-quality community structures can significantly improve the spread of information. However, this improvement is not guaranteed in all scenarios.
As future work, I plan to evaluate the performance of $\alpha$-HCIM under additional diffusion models, such as the Linear Threshold Model and the Heat Diffusion Model.
%
%
%
%
%

%
%
%
%
%
\end{document}